\documentclass[%
 reprint,
 amsmath,amssymb,
 aps,
]{revtex4-2}
\usepackage{color}
\usepackage{graphicx}
\usepackage{dcolumn}
\usepackage{bm}
\usepackage{natbib}

\def\url#1{}

\begin{document}

\preprint{APS/123-QED}

\title{Reprogrammable, in-materia matrix-vector multiplication with floppy modes}

\author{Theophile Louvet}
\author{Parisa Omidvar}%
\author{Marc Serra-Garcia}%
\affiliation{%
AMOLF, Amsterdam
}%

\date{\today}

\begin{abstract}
Matrix-vector multiplications are a fundamental building block of artificial intelligence; this essential role has motivated their implementation in a variety of physical substrates, from memristor crossbar arrays to photonic integrated circuits. Yet their realization in soft-matter intelligent systems remains elusive. Here, we experimentally demonstrate a reprogrammable elastic metamaterial that computes matrix-vector multiplications using floppy modes—deformations with near-zero stored elastic energy. Floppy modes allow us to program complex deformations without being hindered by the natural stiffness of the material; but their practical application is challenging, as their existence depends on global topological properties of the system. To overcome this challenge, we introduce a continuously parameterized unit cell design with well-defined compatibility characteristics. This unit cell is then combined to form arbitrary matrix-vector multiplications that can even be reprogrammed after fabrication. Our results demonstrate that floppy modes can act as key enablers for embodied intelligence, smart MEMS devices and in-sensor edge computing.

\end{abstract}

\maketitle


\section{Introduction}

Linear transformations, encoded in matrix-vector products, are ubiquitous in modern computing; with applications spanning from deep learning to 3D graphics \cite{Alexa2002} and data compression \cite{Kambhatla1997}. In fact, universality theorems prove that any function can be approximated by a network of programmable linear transformations, interconnected by fixed nonlinear activation functions \cite{HORNIK1989359}. This widespread applicability has motivated the development of specialized hardware to compute matrix-vector products, ranging from the well-established Graphics Processing Unit (GPU), to emergent physical computing approaches such as photonic integrated circuits \cite{Ashtiani2022} and memristor crossbar arrays \cite{Yao2020}. 

Within the paradigm of physical computing, mechanics has gathered considerable excitement for a range of special-purpose but highly relevant applications \cite{yasuda2021mechanical}: Mechanics offers the capability to directly process signals such as speech \cite{Dubcek2024} or gait \cite{dion2024sensor} without transduction into the electrical domain---facilitating zero-power, in-sensor information processing; it forms the basis for embodied intelligence in soft robotic structures---enabling such robots to mimic the agility and dexterity of living organisms \cite{Rafsanjani2018, Hu2018, Yasuda2017}, and, when scaled down to microscopic dimensions \cite{Ghadimi2018, Beccari2022}, can perform computations with ultra-low energy consumption \cite{RoukesScience, serra2019turing}---enabling intelligent devices that can operate under extreme power constraints. The overarching challenge in achieving mechanical computation lies in mapping relevant information-processing operations, such as matrix-vector multiplications, into the natural responses of mechanical systems.  

    \begin{figure}[h]
        \centering
        \includegraphics[width=\columnwidth]{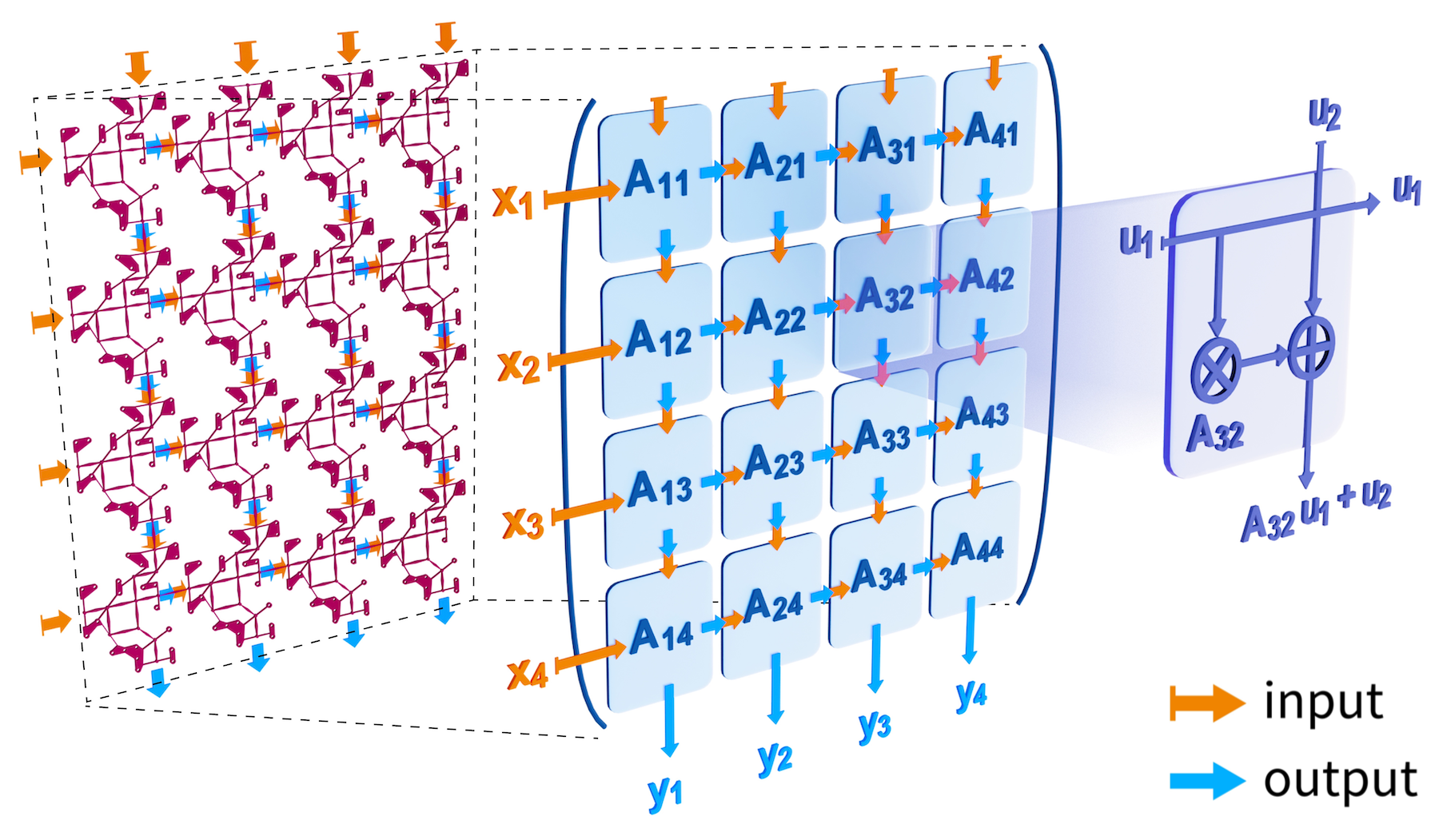}
        \caption{ A metamaterial (red) computes a matrix-vector multiplication by decomposing it into elementary operations (blue), corresponding to individual matrix elements. Each elementary operation (purple) is implemented by a metamaterial unit cell. The input vector is presented as a displacement applied on the input control rods (orange, $x_i$) and the result is deterined by measuring the displacement of the output rods (blue, $y_i$). Each internal output is connected to an input, ensuring that the resulting metamaterial is frustration-free irrespective of the particular matrix being implemented.}
        \label{fig:assembly}
    \end{figure}

In this work, we experimentally demonstrate a soft mechanical metamaterial that computes a matrix-vector multiplication using floppy modes (zero modes). The paper is structured as follows: We start by decomposing the matrix-vector product into a network of elementary operations, each of which will be computed by a single metamaterial unit cell (Fig.~\ref{fig:assembly}). Then, we implement this unit cell in a planar network of point masses subject to ideal constraints. The resulting unit cell presents two zero modes that branch and cross---as required by the matrix-vector multiplication, and preserves deformation compatibility when tiled. Afterwards, we investigate, using Finite Element Method (FEM) simulations, how a real metamaterial geometry deviates from the ideal point-mass model model. We use automatic differentiation \cite{Bordiga2024} to adjust the design and preserve accuracy in the presence of nonzero bending stiffness and finite compliance, and discuss the limitations of the metamaterial arising from such real-world effects. Finally, we experimentally realize the metamaterial using water-jet cut rubber, and characterize its displacement using optical flow. We experimentally characterize the metamaterial unit cell, a 2x2 matrix-vector product, and a tunable unit cell that can be reversibly switched between multiple matrix coefficients. We observe that the metamaterial can compute the desired matrix-vector product with remarkable accuracy, even in the presence of fabrication tolerances and material non-idealities. These results demonstrate that floppy modes can be harnessed to implement intelligent responses in elastic systems.

\section{Design}

Among the deformation responses of mechanical systems, floppy modes, or zero-modes, are unique in the fact that they are associated to a vanishingly low amount of stored elastic energy, i.e., they require small forces to actuate. Thus, they are a prime candidate for computation in passive elastic systems where deformations must be driven by the enegy available in the signal. Floppy modes originate in a rank deficiency of the rigidity matrix, caused by either an insufficient number of constraints \cite{Hu2023, Czaijkowski2024, Maxwell1864, CALLADINE1978}, or by the topological properties of the system \cite{kane2014topological}. Their global nature and topological connections point towards exciting research questions, but also pose a challenging design problem for applications that require a large number of on-demand floppy modes, such as matrix-vector multiplication (a matrix-vector product requires a number of modes equal to the dimension of the input space). Several approaches towards on-demand floppy modes have been introduced: Combinatorial methods draw from a small set of unit cells with well-defined compatibility characteristics \cite{Coulais2016, Coulais2021, Hu2023}. They can be highly efficient, but are limited by the small number of unit cells and large design space.  Techniques based on iterative addition of constraints can accurately produce designs with a few desirable modes, but introduce additional unwanted modes as a by-product  \cite{dykstra2023inverse}. Finally, machine learning methods can identify the number of modes corresponding to a particular unit cell \cite{Mastrigt2022}, but they do not yet possess the capability to solve inverse design problems. While progress has been made in the particular case of line modes \cite{ Lubensky_2015, GUEST2003383, PAPKA1998239, PAPKA19982765, Coulais2021}, branches and crossings---necessary for the realization of mechanical information-processing circuits---still pose a challenge for these methods. We will tackle the challenge of designing a material with a large set of complex floppy modes by introducing a unit cell with well-defined compatibility characteristics, that can be combined to achieve the desired response.

In this section, we introduce a metamaterial design that performs a matrix-vector multiplication $\vec{y}=A\vec{x}$. The input vector $\vec{x}$ is presented to the system by prescribing the displacement of a set of input degrees of freedom, and the output result $\vec{y}$ is read out by measuring the displacement of a different set of degrees of freedom. Since the system should be able to respond to any (small) input vector with no resistance, the space of floppy modes should span all possible input displacement combinations. At the same time, we require the output displacement to be entirely prescribed by the input vector, meaning that the system should not present additional floppy modes. For large input spaces, this involves designing a large number of modes---beyond the limit of direct optimization methods. We can overcome this limitation by decomposing the matrix-vector multiplication as a planar network of computation tiles (Fig 1), whose outputs $u_1$ and $u_2$ are related to the inputs $v_1$, $v_2$ as $v_1=A_{ij}u_1+u_2$ and $v_2 = u_1$. Each tile can be understood as individually implementing a small matrix-vector multiplication, with the form
\begin{equation}
\begin{aligned}
\begin{pmatrix} 
v_1\\
v_2
\end{pmatrix} = 
\begin{pmatrix} 
A_{ij} & 1\\
1 & 0
\end{pmatrix} 
\begin{pmatrix} 
u_1\\
u_2
\end{pmatrix}.
\end{aligned}
\label{eqn:sysexpanded}
\end{equation}
Each individual computational tile is physically implemented by a metamaterial unit cell. Since unit cells are only required to perform a 2$x$2 matrix-vector multiplication, they require only two floppy modes. This simplified requirement can be met with existing floppy mode design methods---here, we solve it by hand-designing the unit cell. 
    \begin{figure}[t]
        \centering
        \includegraphics{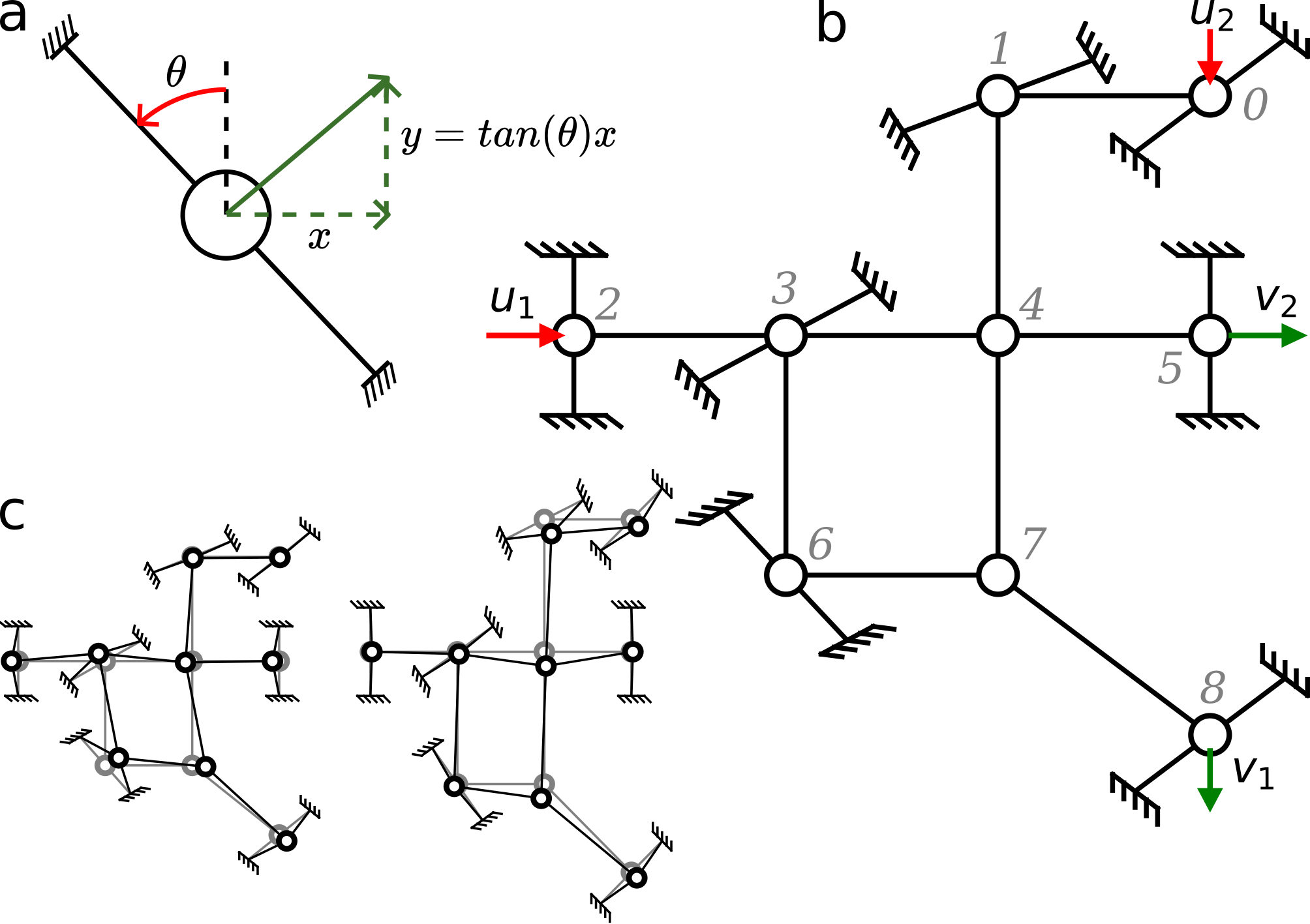}
        \caption{(a) single degree of freedom, the angle of the support beams control the coupling between x and y (b) the unit made of 9 DOFs. The angle at site 3 and 6 (0 and 1) control the coupling between $u_1$ ($u_2$) and $v_1$. (c) the 2 floppy modes of the system}
        \label{fig:unit}
    \end{figure}

By construction, the tiling in Fig. \ref{fig:assembly} exhibits the correct number of global floppy modes: Each individual row added to the system will add $2n_{col}$ floppy modes, but will also incorporate $2n_{col}-1$ constraints ($n_{col}$ constraining the input $u_2$ with the output $v_1$ of the upper row, and $n_{col}-1$ intra-row constraints). This will increase the number of floppy modes by one, reflecting the fact that the extended system contains an extra input. Adding an additional column will add $2n_{row}$ floppy modes, but also $2n_{row}$ constraints ($n_{row}$ constraining the input $u_1$ of each site with the output $v_2$ of the previous column, $n_{row}-1$ intra-column constraints plus an extra constraint setting the input $u_2$ of the topmost site to zero, as shown in Figure \ref{fig:assembly}). Since the constraints always connect an output (by definition prescribed) with an input (by definition free), and the directed computation graph presents no cycles, the constraints will neither neither be redundant nor introduce frustration. Such designer compatibility relation is reminiscent of traditional combinatorial design approaches for floppy mode systems \cite{Coulais2016}. However, here the site is continuously parameterized by $A_{ij}$, thus, the same design can encode a variety of matrices by designing the individual building blocks. This is a key to our ability to design and re-program the metamaterial.

    \begin{figure}
        \centering
        \includegraphics{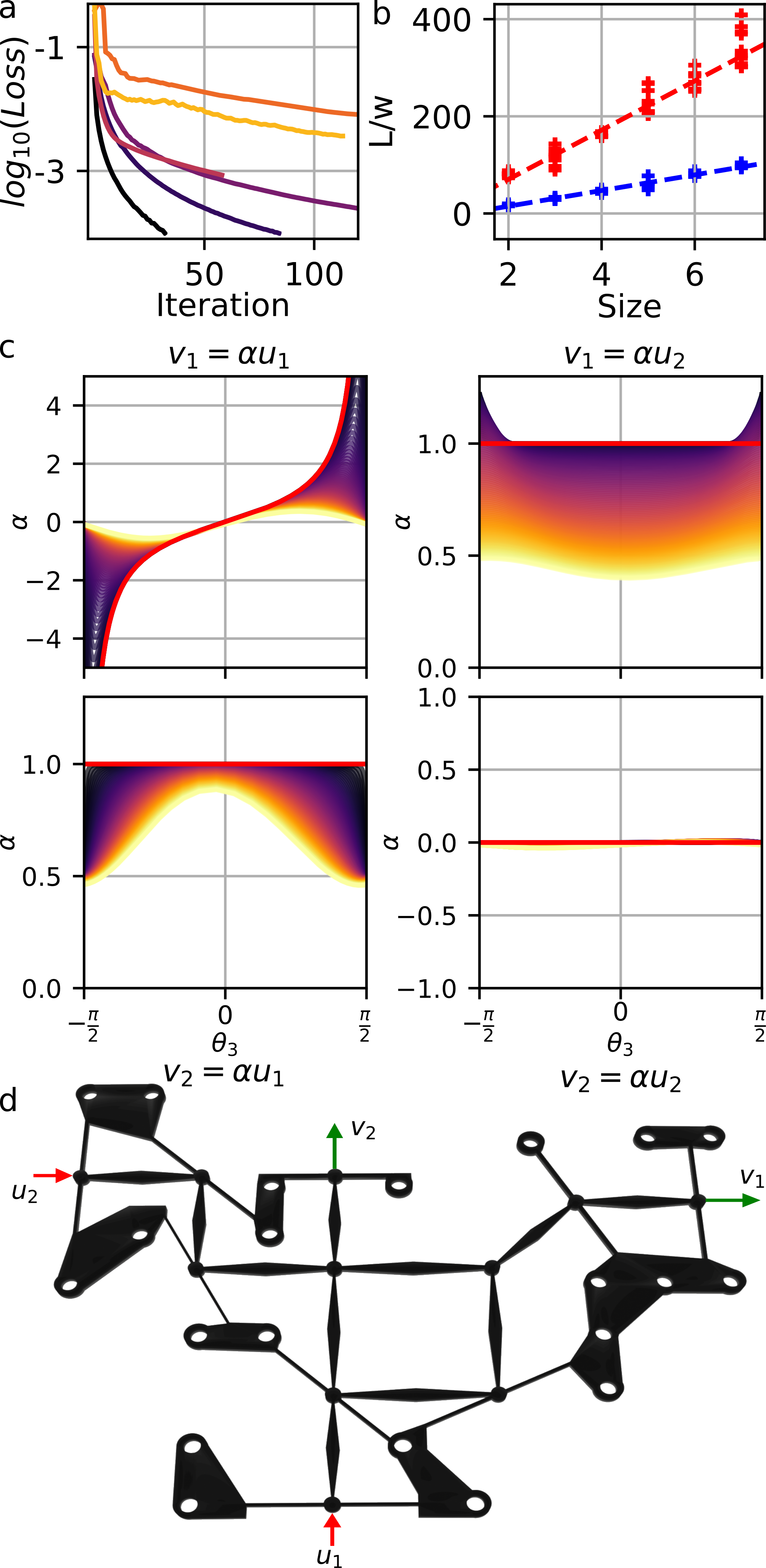}
        \caption{(a) Evolution of the loss function during the optimization of the design from 1x1 lattices (black) to 6x6 (yellow). For large matrix sizes, the loss function cannot go below a saturation threshold.  (b) Minimum aspect ratio required to accurately approximate a set of unitary random matrices. In the red curve, only $\theta_3$ is varied, while for the red curve, 5 angles are allowed to vary. (c) Input/output relation for a single unit cell, as a function of the angle of site 3, from a low (black) to high (yellow) amount of bending stiffness (black). We see that as the the aspect ratio is reduced, both the response at each given angle, and the maximum attainable matrix coefficient are lowered. (d) 3D model of a unit cell}
        \label{fig:bending}
    \end{figure}

The unit cell design consists of nine sites (Fig. \ref{fig:unit}a,b) that can move on a plane. Before introducing any constraints, the sites are free to move along both $x$ and $y$ axes, and thus the constraint-free unit cell has 18 floppy modes. As a consequence of the Maxwell-Calladine theorem, restricting the unit cell to the two prescriptive modes requires incorporating 16 linearly-independent constraints. To produce an initial guess for the metamaterial design, we start by modelling the constraints as ideal rods, for which bending stiffness is zero; and we consider the displacements infinitesimal, so as not to alter the length of the rods. We incorporate two classes of constraints: Local constraints (Fig. \ref{fig:unit}a)---represented by a pair of rods connected to a fixed boundary conditions, which restrict the site motion to a specific line along the $x$ and $y$ axes---introducing a rank-1 condition of the form $(\cos \theta, \sin\theta)^T(x,y)=0$; and pairwise constraints, which are realized by rods connecting two sites, resulting in a rank-1 condition of the form $(\cos \theta, \sin\theta)^T[(x_1,y_1)-(x_2,y_2)]=0$ between the displacements of the two sites. 
  
The proposed unit cell design is introduced in Fig. \ref{fig:unit}b. The two floppy modes (Fig. \ref{fig:unit}c), spanning the two-dimensional input space, can be unterstood as follows: When the input $u_1$ is displaced, the rank-1 conditions introduced by the couplings between sites 2-3, 3-4 and 4-5 cause the output degree of freedom $v_2$ (horizontal displacement of site 5) to follow the input degree of freedom (satisfying the condition $v_2=u_1$ required by Eq. \ref{eqn:sysexpanded}). At the same time, the condition introduced by the local constraints in site 3 couple its vertical displacement to its horizontal displacement, with a proportionality factor given by $\tan \theta_3$. This vertical displacement is then transferred to site 6, and converted to a horizontal displacement, scaled by the factor $\tan^{-1} \theta_6$, and transferred to the horizontal displacement of site 7. Similarly, the displacement $u_2$ is scaled by $\tan^{-1} \theta_0$ and $\tan \theta_1$, and then transferred through sites 1-4 and 4-7 to the vertical displacement of site 7. The diagonal coupling between sites 7 and 8 ensures that the displacement along the diagonal direction of site 8 matches the diagonal displacement of site 7. This diagonal displacement the projection on $1/\sqrt{2}$($dx_7$, $dy_7$), where $dx_7$ is scaled displacement of input $u_1$, and $dy_7$ is the scaled displacement of input $u_2$. This projection amounts to adding the horizontal and vertical displacements together. A local constraint on site 8 prevents the output site to move in the tangent direction to the 7-8 couplings. The output vertical displacement amounts to the projection of the diagonal displacement into the y-axis, introducing a total scaling of $1/2$. By setting the angles $\theta_0$ and $\theta_1$ to compensate for the factor of $\frac{1}{2}$, and setting the angles $\theta_3$ and $\theta_6$ to implement a coefficient $2A_{ij}$, the resulting system satisfies the required input-output characteristics.

The final system retains two floppy after the addition of the 16 constraints (7 local constraints plus 9 coupling constraints). These two resulting floppy modes are plotted in Fig. \ref{fig:unit}c. Since any prescribed input displacement can be expressed as a combination of these  modes, in the ideal rod approximation, this (idealized) matrix-vector product requires no energy to be actuated. The floppy modes in our system are reminiscent to line-modes \cite{ Lubensky_2015, GUEST2003383, PAPKA1998239, PAPKA19982765, Coulais2021} observed in two-dimensional soft matter systems. However, in contrast with prior works, our unit cell design allows for line modes to branch and cross, as required for the realization of the matrix-vector product.

    \section{Numerical validation}

    Until now, we have considered a design composed of ideal rods (zero bending stiffness) under infinitesimal displacements. However, these assumptions do not accurately describe the dynamics of experimental samples. To capture the effect of the finite bending stiffness of the rods, we construct a three-dimensional finite element simulation of the metamaterial (see Appendix A). Using automatic differentiation, it is also possible to compensate for some of these errors by fine-tuning the geometry.
    
    Attempting to compensate for real-beam effects by adjusting the beam angles illustrates the limitations of the system (Fig. \ref{fig:bending}a): For small matrices, we can approximate the objective transformation with arbitrarily high accuracy. However, as the matrix size grows beyond $4x4$, the optimization algorithm cannot decrease the loss function below a given floor value. The origin of this error can be found in the relation between beam angle $\theta_3$ and matrix coefficients (Fig. \ref{fig:bending}c). While systems composed of ideal rods (infinite aspect ratio/zero bending stiffness) can approximate any matrix coefficient, rods with finite aspect ratio can only achieve a maximum matrix coefficient. This limitation arises from the ratio between longitudinal and bending stiffness in finite aspect ratio rods: Increasing rod angles has the effect of stiffening the input of the unit cell. Beyond a specific angle, the decrease in input displacement due to the finite longitudinal stiffness of the preceding rods exceeds the increase in output displacement due to the increase in angle. At this point, the matrix coefficient has reached saturation. We quantify this effect by determining the maximum matrix size that can be represented using a given aspect ratio. To do so, we consider 5 random matrices of each size, and perform an optimization to determine the minimum aspect ratio at which it is possible to achieve arbitrary accuracy. The results are plotted in Fig. \ref{fig:bending}b)---highlighting a linear relation between aspect ratio and maximum matrix size. Given that MEMS resonators with aspect ratios up to 1000:1 \cite{glauvitz2014mems} are possible, we anticipate that our results can lead to matrix-vector products with dimensions above 64x64.

    \section{Experiments}

    \begin{figure}
        \centering
        \includegraphics{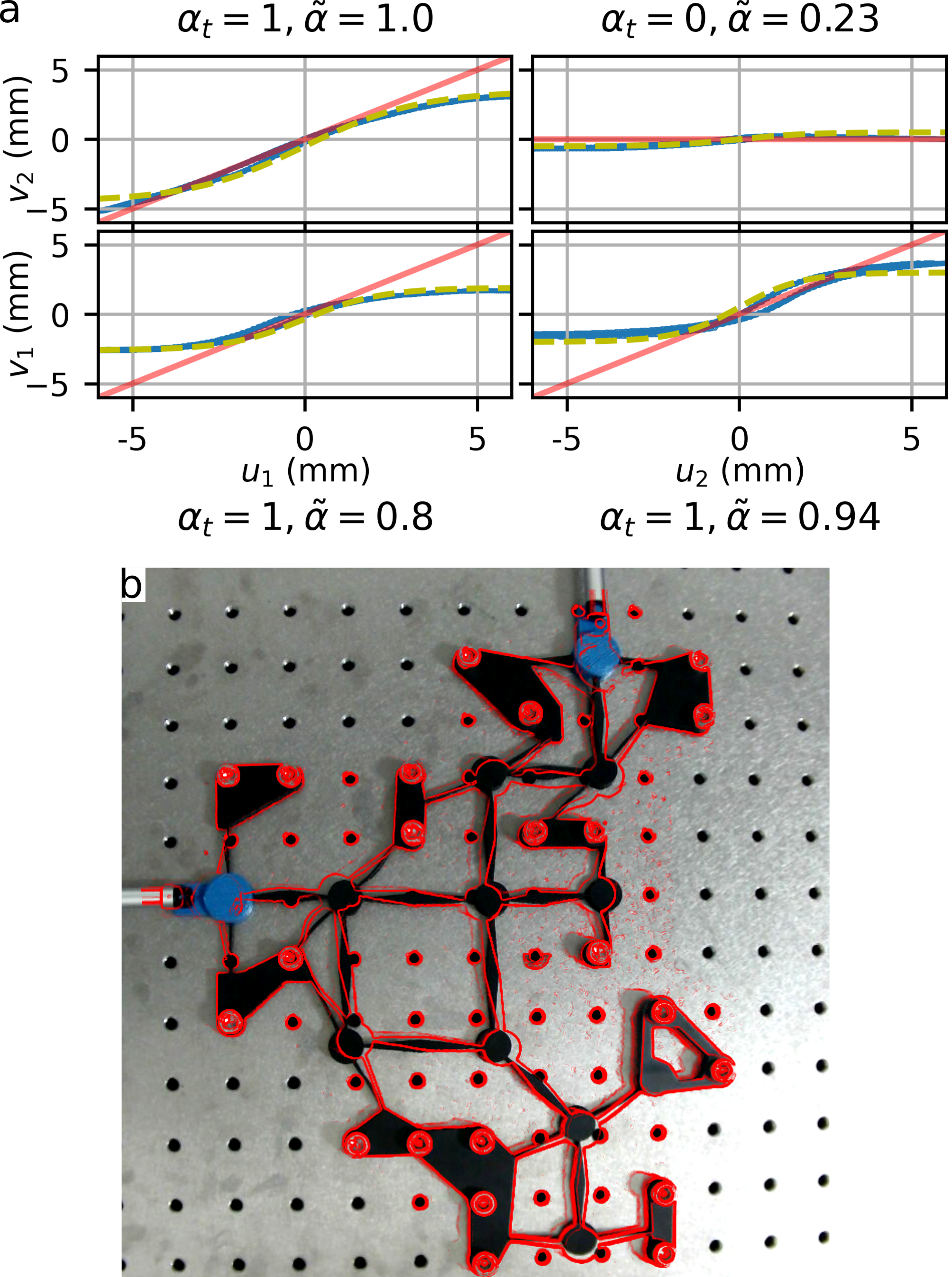}
        \caption{(a) Complete experimental input/output transfer relation for a single unit cell, the measurements, in blue, are gathered from an optical flow algorithm. The fitted linear response (solid red lines) is within 20$\%$ of the target matrix value. The dashed olive line corresponds to a sigmoidal fit of the output displacement.  (b) Picture of the experimental setup during characterization. The blue components are the output of the linear actuators.  Videos of the experiment are provided as supplementary information. The red lines represent the deformation of the unit, it is computed using an edge detection algorithm and some edited to remove the edges of holes around the sample.}
        \label{fig:1x1}
    \end{figure} 

    \begin{figure}[b]
        \centering
        \includegraphics{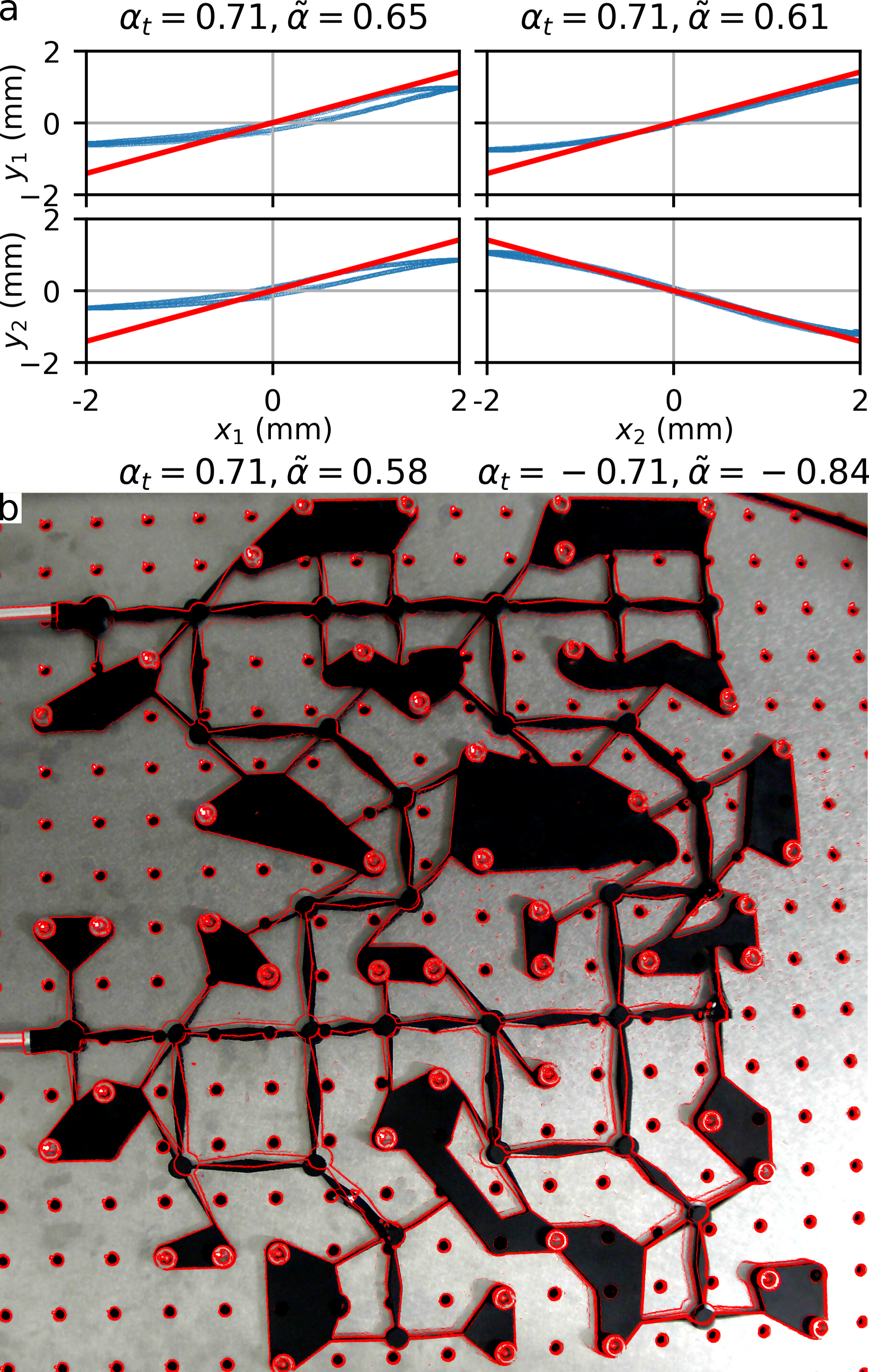}
        \caption{(a) input/output relationship of a 2x2 lattice. The matrix of the transformation is within 15\% of the target matrix (b) Picture of the 2x2 lattice, with the deformation plotted in red. The first output is zero because the components cancel each other out, while they add for the second output. }
        \label{fig:2x2}
    \end{figure}
    
    We fabricate and optimize the design on a rubber sheet with a thickness of $6\,\text{mm}$, using water-jet cutting. The input displacement is prescribed using stepper motors, and the output displacement is measured with a camera and processed using an optical flow algorithm. The experiments on a single unit cell reveal that, in the linear regime (where input displacements remain below a few mm), the displacement transfer matrix coefficients between input and output are within 20\% of the target value from the ideal unit. Beyond the linear regime (around 2 mm---for a unit of 250x275mm), the output displacement saturates. Interestingly, the saturation function resembles a sigmoid function. In neural networks, sigmoid functions are one of the most commonly used nonlinear activation functions---thus, the saturation characteristics of the material present an exciting prospect for the realization of full machine learning models in elastic systems. In addition to the sigmoidal-like saturation, the experimental results present two deviations from the ideal response: First, we observe that the response presents some hysteresis---which we attribute to the visoelastic response of the polymer material and non-idealities in the actuator-sample couplings. Second, the linear region is asymmetric---saturation occurs earlier under compressive loads than under tensile loads. Experiments on a metamaterial consisting of two unit cells reveal a similar response---with $12\%$ error between the objective matrix and the measured linear transformation, and similar hysteresis and saturation characteristics.
   \\

    \begin{figure}
        \centering
        \includegraphics{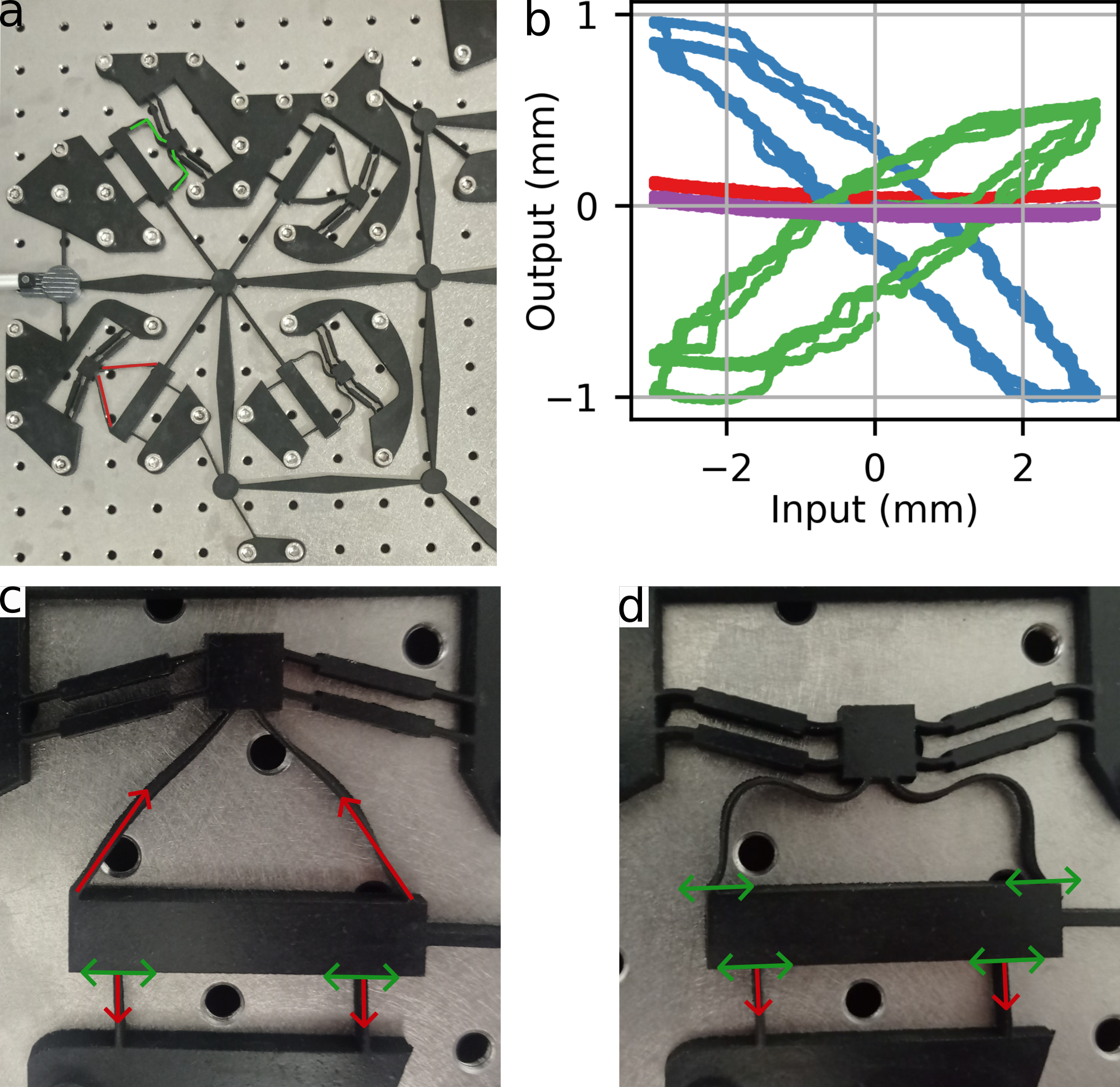}
       \caption{(a) Picture of the sample during characterization, with the lines showing a compliant mechanism in the low-stiffness (green) and high-stiffness (red) configurations. In the low-stiffness configuration, the beams are buckled, while in the the high-stiffness configuration, the beams are straight. (b) Experimental results demonstrating the switching between three different matrix coefficients, $A_{ij}= +.25$ (green), $A_{ij}= 0$ (purple, red),  $A_{ij}= -.33$ (blue). Switching between green and blue curves does not alter the number of floppy modes. (c, d) Variable stiffness mechanism in the clamped position (c), where the tension in the upper beams hold the shaft in place and (d) in the 'free' position the buckling beams allow the shaft to move}
        \label{fig:tunable}
    \end{figure} 
    

We now proceed to demonstrate a programmable metamaterial that can be re-configured to implement different matrix-vector products. Reconfigurable floppy modes are an timely area of research, with existing results focusing on controlling the number of modes \cite{Lei2023, Wu2024}. The requirements for a programmable matrix-vector product are different, as the number of modes shall remain constant while the mode shape is altered. To allow for multiple possible values for the transformation coefficients, we incorporate additional rods in the site 3, the site that defines $A_{ij}$, the coordinate transformation. Increasing the number of constraining rods causes the system to be over-constrained---thus eliminating the zero modes. To achieve programmability, only a sub-set of the rods is active in each configuration. We accomplish this by incorporating bi-stable elements into the metamaterial \cite{chen2021reprogrammable}. In particular, we use a variable stiffness compliant mechanism \cite{Kuppens2021}, connected to the base of each of the support rods (Fig. \ref{fig:tunable}a). When the compliant mechanism is in the low-stiffness configuration, the rods are free to move, and the constraint does not act into the system. By switching the compliant mechanisms between different configurations, the unit cell coefficient $A_{ij}$ can be switched between three different values (Fig. \ref{fig:tunable}b)---although only two of them present the correct number of floppy modes. The programmable unit cell shows an increased hysteresis due to slack in the fabricated compliant mechanism. There are several avenues to increase the resolution in the matrix coefficients, including using more than four switchable support rods, or cascading multiple unit cells by cascading programmable units.

    \section{Discussion}

  The present work demonstrated a mechanical metamaterial capable of re-programmable matrix-vector multiplication. The results can be interpreted in the light of the in-materia computing paradigm, where non-traditional material responses---here, floppy modes---are harnessed to perform computations---in this case, matrix-vector products. The key factor limiting the matrix size is the aspect ratio of the supporting beams, with matrix sizes above 64x64 attainable within present technology. Since, relevant machine learning such as phoneme recognition can be efficiently encoded in classifiers acting on a as little as 39 features \cite{salomon2001support}, the proposed passive matrix-vector multiplier can have an enabling role towards passive speech-recognizing MEMS \cite{Dubcek2024}. We have also shown a reprogrammable unit cell where every weight $A_{ij}$ can be adjusted from a pair of values while preserving the number of zero modes. This level of programmability---that can be expanded using more complex designs---has the potential to be applied to relevant problems, as machine learning models are known to be resilient to extreme weight quantization, down to 3 values per weight \cite{ma2024era}.

\nocite{*}

\bibliography{refs}

\section{Appendix}

\subsection{Simulation and optimization}
\label{subsec:simulation}

\begin{figure}[h]
    \centering
    \includegraphics{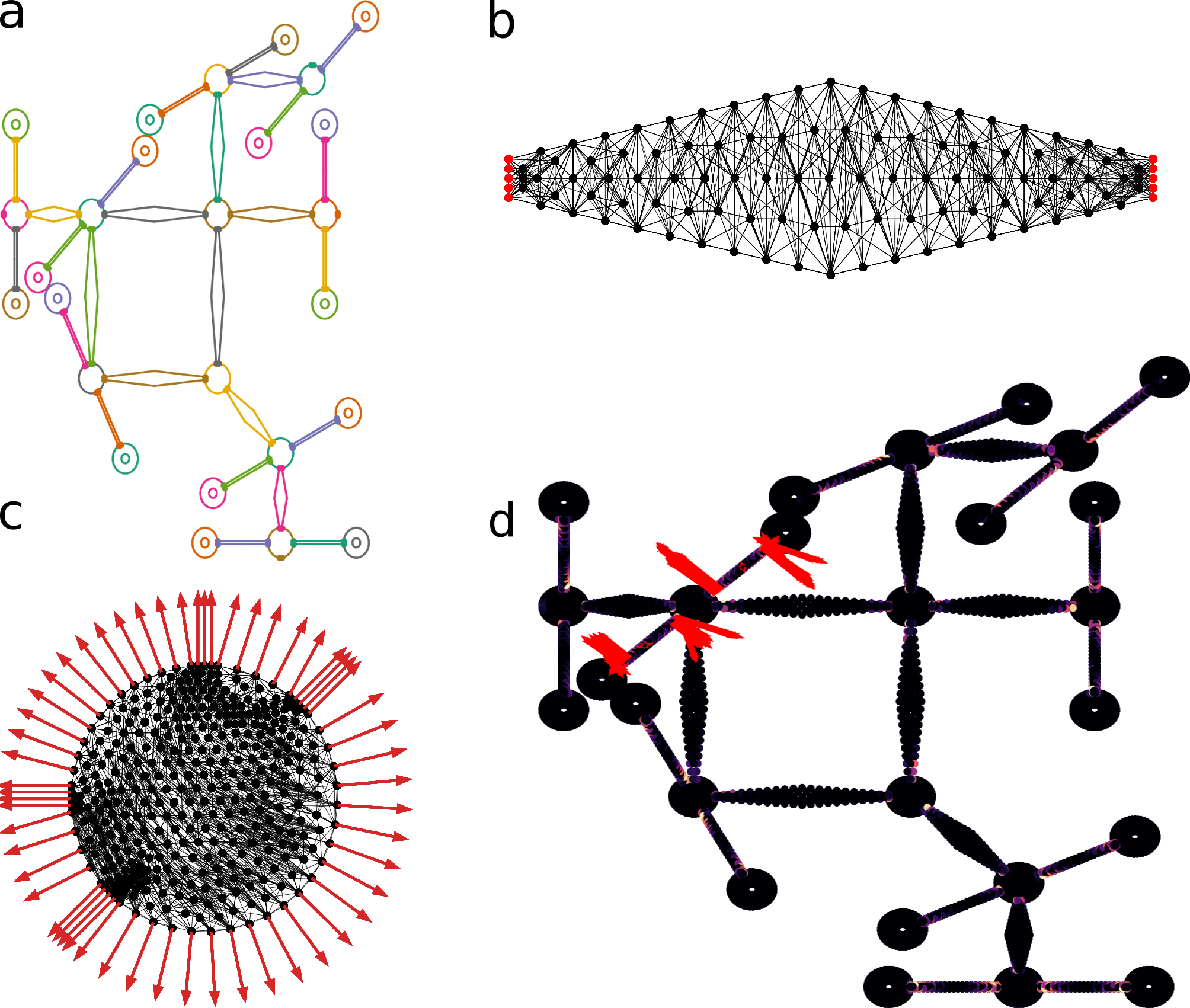}
    \caption{(a) Division of the planar geometry into sub-components (each component is plotted in a different color). The solid dots represent the contact points---that are automatically matched between each component to ensure compatibility.  (b) finite element meshing of a single component, red dots represent boundary degrees of freedom, which match between components, which are deterministically positioned and thus guaranteed to match between components. (c) Gradient of the external mesh node coordinates with respect to a high-level parameter (in this case, the radius) (d) Gradient of the misfit $\eta$ with respect to the mesh node coordinates. }
    \label{fig:supp}
\end{figure}

We simulate and optimize the metamaterial using the finite element package FEniCSx \cite{Scroggs2022, 10.1145/2566630, baratta_2023_10447666, 10.1145/3524456}. To do so, we partition the structure into sub-components \ref{fig:supp}. Each sub-component is obtained by extruding a two-dimensional geometry, composed of lines and circle arcs. The lines and arcs are generated using Python functions, from a set of high-level parameters, such as centers, angles, radiuses and thicknesses. We use JAX to map the high-level parameters into the elementary features (arcs and lines) to allow for  automatic differentiation. Automatic differentiation is used to assemble the geometry from a graph of interconnected components. The graph specifies the topological characteristics of the system---i.e., which component is connected to which. Then, by computing the Jacobian of the distances between component boundary contact point with respect to the high-level parameters, $J_{contact}$, a distance minimization algorithm is capable of automatically constructing the full metamaterial geometry---enforcing continuity between components. \\

To simulate the system, we extract the stiffness matrix for every component using second-order Lagrange elements and a linear elastic energy function. Since we are only interested in the degrees of freedom located on boundaries; we use the Guyan reduction \cite{Guyan1965}. This computation allows us to determine the exact, statically condensed component stiffness matrix, that contains only a few hundred rows and columns. The determination of the statically condensed stiffness matrix, requiring a matrix inversion, is performed on the GPU---we found out that, given the relatively small size of each component (a few thousands of degrees of freedom), GPU-based dense linear algebra achieved the highest performance. The matrices for each component can be calculated fully in parallel.

To be able to assemble a full system matrix from the reduced component matrices, we make sure that all connecting boundaries are meshed with identical, regular grids. Then, we assemble the statically condensed stiffness matrix from every component into a global stiffness matrix. This global stiffness matrix can also be decomposed in inner degrees of freedom (corresponding to components that are connected to each other (not accessible from outside the metamaterial), and external boundaries (that correspond to the system inputs and outputs). Since we are interested in the response of the metamaterial to displacements in the external boundaries, we apply a second Guyan reduction. This second Guyan reduction can be very expensive for large systems. Thus, it is performed recursively, assembling intermediate structures with a larger and larger number of components until the full system is assembled.\\

Once the global stiffness matrix has been statically condensed to remove internal boundaries, we divide its degrees of freedom into input and output displacements,
\begin{equation}
K_{global}(\vec{\theta}) = \begin{pmatrix}
K_{in,in}(\vec{\theta}) & K_{in,out}(\vec{\theta}) \\
K_{in,out}^T(\vec{\theta}) & K_{out,out}(\vec{\theta}),
\end{pmatrix}
\label{eq:inpout}
\end{equation} where $\vec{\theta}$ is the vector of high-level geometric parameters.

The transfer function between input and output displacements can be calculated as $T(\vec{\theta}) = -K_{out,out}^{-1}(\vec{\theta})K_{in,out}^T(\vec{\theta})$. Once this transfer matrix $T(\vec{\theta})$ has been calculated, we prescribe the inputs to move along a specified direction---given by the displacement direction of the actuators, and project the resulting output into the direction of measurement. The projection $A(\vec{\theta})=\hat P_{out}T(\vec{\theta})P_{in}$, where $P_{in}$ and $P_{out}$ are the input and output projectors, provides the effective implemented matrix $A$. In our case, $A$ should match the objective matrix of the design, $T$. The agreement between the implemented and objective matrices can be quantified by defining a loss function of the form\

\begin{equation}
\eta(\theta) = \sum_{i=0}^{N_{in}}\sum_{j=0}^{N_{out}}(T-A(\theta))^2
\label{eq:lossfunction}
\end{equation}

The entire evaluation tree, from the component stiffness matrices to the loss function $\eta(\theta)$ can be automatically differentiated using JAX. Once the change in misfit with respect to the component stiffness matrices has been determined, the gradient with respect to the finite element mesh can be calculated in FEniCSx using the built-in automatic differentiation capabilities in the UFL library \cite{Ham2019}. This produces a gradient of the misfit with respect to the mesh coordinates (Fig. \ref{fig:supp}d), indicating how the mesh should be deformed to improve the matrix-vector multiplication accuracy. This mesh gradient can be projected into a gradient of the boundary coordinates with respect to the high-level parameters (Fig. \ref{fig:supp}d), producing the gradient of the misfit with respect to the high-level parameters. It should be noted that not all parameter updates are valid; some changes to the geometry will result in disconnected components. To prevent this, we project the gradient into the nullspace of the contact point distances $J_{contact}$ with respect to the high-level parameters, and perform an additional boundary distance minimization optimization to preserve the well-connectedness of the metamaterial.






\end{document}